%% file: usenixsecurity2026.tex
\documentclass[letterpaper,twocolumn,10pt]{article}
\usepackage{usenix}
\usepackage[margin=1in]{geometry}
\usepackage{array}
\usepackage{tabularx}
\usepackage{multirow}
\usepackage{makecell}
\usepackage[table]{xcolor}
\usepackage{rotating}
\usepackage{ragged2e}
\usepackage{enumitem}
\usepackage{amssymb}
\usepackage[table]{xcolor}
\usepackage{booktabs}
\usepackage{tabularx}
\usepackage{makecell}
\usepackage{tabularray}
\usepackage{xcolor}
\usepackage{ulem}
\usepackage{pifont}
\usepackage{tikz}
\usepackage[most]{tcolorbox}
\usepackage{xcolor}
\usepackage{colortbl}
\usepackage{booktabs}
\usepackage{fontawesome5} 
\usepackage{placeins}
\usepackage{float}
\usepackage{colortbl}
\usepackage[table]{xcolor}
\usepackage{tabularray}
\usepackage{graphicx}
\usepackage{caption}
\usepackage{tabularx}
\usepackage{multirow}

\usepackage{xurl} 
\usepackage{pifont}

\definecolor{HeaderBlue}{RGB}{225,235,248}
\definecolor{RowGray}{RGB}{246,248,250}
\definecolor{Good}{RGB}{34,139,34}
\definecolor{Bad}{RGB}{178,34,34}
\definecolor{Mid}{RGB}{184,134,11}

\newcommand{\YesIcon}{{\color{Good}\faCheckCircle}}
\newcommand{\NoIcon}{{\color{Bad}\faTimesCircle}}
\newcommand{\PartIcon}{{\color{Mid}\faAdjust}} 

\definecolor{HeaderBlue}{HTML}{1F4E79}
\definecolor{RowGray}{HTML}{F5F7FA}
\definecolor{Good}{HTML}{2E7D32}
\definecolor{Warn}{HTML}{F9A825}
\definecolor{Bad}{HTML}{C62828}
\definecolor{GoodBg}{HTML}{E8F5E9}
\definecolor{WarnBg}{HTML}{FFF8E1}
\definecolor{BadBg}{HTML}{FFEBEE}

\renewcommand{\arraystretch}{1.2}

\newtcolorbox{infobox}{
  colback=blue!5,        
  colframe=blue!60!black,
  arc=6pt,               
  boxrule=0.8pt,         
  left=8pt,
  right=8pt,
  top=6pt,
  bottom=6pt
}
\usepackage{mdframed}
\usepackage{xcolor}
\usepackage{xcolor}
\newcommand{\cmark}{\ding{51}}
\definecolor{lightblue}{RGB}{230,240,255}
\usepackage{tikz}
\usepackage{amsmath}

\newcolumntype{L}[1]{>{\RaggedRight\arraybackslash}p{#1}}
\newcolumntype{Y}{>{\RaggedRight\arraybackslash}X}

\definecolor{openchg}{HTML}{CFEBD6}      
\definecolor{selfenh}{HTML}{D5E7FB}      
\definecolor{conserv}{HTML}{F9E6C7}      
\definecolor{selftrans}{HTML}{F8D1D6}    
\definecolor{interaction}{HTML}{D9D2F3}  
\definecolor{headergray}{HTML}{F2F2F2}

\usepackage[most]{tcolorbox}
\usepackage{xcolor}
\usepackage{newtxtext} 
\usepackage[margin=1in]{geometry}

\definecolor{calloutBack}{HTML}{D7E6F9}  
\definecolor{calloutFrame}{HTML}{A9C3EA} 
\definecolor{calloutBar}{HTML}{2F6FB1}   
\newtcolorbox{mentalcallout}{
  enhanced,
  boxrule=0.6pt,
  colback=calloutBack,
  colframe=calloutFrame,
  arc=4pt,
  left=10pt, right=10pt, top=8pt, bottom=8pt,
  borderline west={4pt}{0pt}{calloutBar}, 
}

\usepackage{filecontents}

\begin{document}

\date{}

\title{\Large \bf V.O.I.C.E (Voice, Ownership, Identity, Control, Expression): Risk Taxonomy of Synthetic Voice Generation From Empirical Data}

\author{
{\rm  Tanusree Sharma$^{1}$ \quad
Anish Krishnagiri$^{1}$ \quad
Lili Dudas$^{1}$ \quad
Ahmed Adnan$^{3}$ \quad
Visar Berisha$^{2}$}\\
{\rm $^{1}$Penn State University \quad
$^{2}$Arizona State University} \quad
$^{3}$ University of Dhaka}


\maketitle

\begin{abstract}
As generative voice models are rapidly advancing in both capabilities and public utilization, the unconsented collection, reuse, and synthesis of voice data are introducing new classes of privacy, security and governance risk that are poorly captured by existing, largely uniform threat models. To fill the gap, we present V.O.I.C.E, a taxonomy of voice generation risk grounded in a multi-source threat modeling effort with 569 incidents from major AI incident database, FTC and Internet Crime Complaint Center (IC3); 1067 direct incident reports from U.S. based participants across diverse groups (including voice actors, internet personalities, political personnel, and general public); and 2,221 Reddit discussions. Grounded in real-world data, our taxonomy explicitly models how risk emerges, interact with contextual factors such as degree of  exposure, social visibility, and the availability of legal protections for various affected groups. 

\end{abstract}

\input{sections/1-intro}
\input{sections/2-background}
\input{sections/2-regulation-analysis}

\input{sections/3-method}

\input{sections/4-results}

\input{sections/5-discussion}
{\footnotesize\bibliographystyle{plain}
\bibliography{sample}} 



\appendix
\input{sections/Appendix}



\end{document}

%% file: sections/1-Intro.tex
\section{Introduction}
\label{sec:intro}
Voice constitutes one of the most ubiquitous modes of human-AI interaction, yet it is simultaneously biometric, uniquely identifying, and deeply embedded in social and institutional contexts. Rapid advances in generative voice technologies expanded opportunities for creativity, accessibility, and automation~\cite{garofolo1993timit, song2024tacolm, elevenlabs, sharma2025before, sharma2023disability}. At the same time, these developments have introduced new forms of risk that extend beyond traditional cybersecurity or privacy concerns~\cite{warren2024better, muller2022human} which are evolving faster than the preparedness of policymakers, platform operators, and affected communities. Recent incidents, such as a high school athletic director using AI voice cloning to fabricate racist statements attributed to a principal~\cite{apnews_ai_principal_audio_2024}, and criminals using synthesized executive voices to authorize fraudulent wire transfers exceeding \$35 million ~\cite{muller2022does}, highlight the severe consequences when emerging threats go unrecognized.

Recent work analyzing 37.9 hours of in-the-wild recordings from celebrities and politicians found nearly half (17.2) hours are deepfakes, highlighting a troubling reality that many detection methods fail to generalize to real-world conditions~\cite{muller2022does, berisha2025speech}. As AI detection evolves into a cat-and-mouse game, recent work shows that relying solely on detection based system is insufficient~\cite{doan2025defending, zhang2025safespeech, yan2025voicewukong}. 
Beyond technical limitations, historically, the adverse lived experiences and harms often emerge from the operation in the world through the interplay of systems and societal power dynamics rather than from technology alone~\cite{shelby2023sociotechnical}. Consequently, even as defenses against voice fraud, scams, and deepfakes improve, individuals’ actual exposure to risk depends on how they engage with these systems within their social environments.

These risks are not distributed evenly. Previous work identified that high-risk groups such as, public figures, politicians, media personalities, social media influencers, and voice actors, often have widespread, easily accessible voice recordings online~\cite{wang2019asvspoof, notMyVoice}. Thus, they encounter distinct categories of harm, including system-bound security risks (voice as biometrics), social system risks (malicious collection and synthesis), unintentional harms (dataset repurposing), and misattribution (non-consensual use for exploitative content including CSAM)~\cite{farid2022deepfakes}.  
However, these attack surfaces may not affect all group equally. While high-profile celebrities like Scarlett Johansson may benefit from legal recourse and platform cooperation, lesser-known general public, voice actors and internet influencers may face disproportionate risks~\cite{kshetri2006simple, sharma2025prac3}, since attackers are incentivized to target individuals with high utility and low resistance, those with weaker legal and platform protection. Despite this asymmetry, existing research has focused primarily on voice-based threats for average end-users and institutions~\cite{almutairi2022review}, there is little we know about how these threats affect individuals whose voices are widely exposed online, even though prior studies suggest such public visibility increases susceptibility to identity-based attacks~\cite{samermit2023millions, han2024pressprotect, sharma2025verifying}. 
Without a comprehensive taxonomy that accounts for these differential vulnerabilities, critical risks may go unrecognized. Thus, we argue for the development of a comprehensive risk framework, accounting for sociotechnical, legal, and platform-specific realities of voice generation based risks. In this paper, we addressed the following research questions.

\textbf{RQ1: Systemization of Risks.} What risks are encountered and anticipated by different affected groups in generative voice technologies? 

\textbf{RQ2: Contextual Risk Factors \& Propagation} How do these risks emerge and compound harm through different interaction pathways with varying availability of regulatory resources?




\textbf{Main Findings.} In this paper, we provide empirical evidence illustrating how people encounter and experience potential risk from voice generation and synthesis.

  \textcolor{green!60!black}{\cmark}
  We developed a multi-stakeholder–centered voice synthesis risk taxonomy (V.O.I.C.E) grounded in real-world data sources, 569 curated incidents, 2,221 Reddit discussions, and 1,067 direct incident reports.

\textcolor{green!60!black}{\cmark}
Our analysis reveals \textbf{six} high-level risk categories: Privacy, Safety \& Data Protection; Authentication, Cybersecurity \& Espionage; Information Integrity \& Authenticity; Individual Rights, Labor \& Commercial Integrity; Platform Governance; and Psychological \& Social Harm each containing multiple medium-level subcategories, totaling 82 distinct low-level risks, as detailed in Figure~\ref{fig:risk}.


\textcolor{green!60!black}{\cmark}
  We analyzed existing U.S. and international regulations to demonstrate that current regulatory resources remain fragmented and insufficient, particularly for low-resource groups with limited legal recourse.

Our findings motivate our discussion of anticipatory and contextual threat modeling, exposure-weighted safeguards, and tiered governance mechanisms that better reflect the sociotechnical realities of voice synthesis risks

%% file: sections/2-background.tex
\section{Background \& Related Work}
\label{lit}

\subsection{History of Audio Dataset in AI Models Innovation}
Audio datasets have served as the cornerstone for speech innovation \cite{garofolo1993timit}. The history of voice technology has undergone a shift in how data is collected, processed, and analyzed, especially in the emergence of advanced AI models~\cite{kahn2020libri, nagrani2017voxceleb}. What began as a small-scale, carefully controlled collection of studio recordings has transformed into harvesting millions of voices directly from the internet to train the AI models for speech synthesis and analysis \cite{kearns2014librivox, panayotov2015librispeech, nagrani2017voxceleb}. For instance, this evolution is evident in the transition from the phonetically balanced small-scale TIMIT studio speech dataset \cite{garofolo1993timit} to the large-scale LibriSpeech audiobook dataset \cite{panayotov2015librispeech, wang2023neural}, and finally to VoxCeleb \cite{nagrani2017voxceleb}, a collection of more than a million audio clips from celebrities scraped directly from YouTube. This exponential growth in data volume has happened with the purpose to train deep learning, and generative AI models effectively, as they require a large amount of data with diversity to generalize successfully, and perform tasks such as speech recognition and zero-shot voice synthesis \cite{agnew2024sound, radford2023robust, wang2023neural}. In recent times, this trend has led to the release of large-scale, multi-domain audio datasets, such as Libri-Light \cite{kahn2020libri} and GigaSpeech \cite{chen2021gigaspeech}. These datasets contain tens of thousands of hours of semi-supervised data for training speech/audio synthesis, conversion, and cloning. While early large-scale  datasets were built with hundreds of individual contributors whose voices were instrumental in speech technologies, including audiobooks and voice assistants. Yet, A decade later, these same contributions now expose people to new risks.
Despite substantial progress in detection systems~\cite{doan2025defending, zhang2025safespeech, yan2025voicewukong}, real-world incidents demonstrate that voice generation continue to affect diverse populations~\cite{warren2024better, muller2022human}.

\subsection{Tools and Algorithms in Voice Generation \& Synthesis}
Early research in Text-to-Speech (TTS) synthesis focused on improving statistical prosody modeling, such as, using acoustic cues for phrase boundary detection\cite{prahallad2010semi}. With the rise of deep learning, the field shifted toward the development of end-to-end architectures for scalable speech synthesis. For example, speaker verification embeddings  multispeaker synthesis \cite{jia2018transfer}, while model like VITS integrated autoencoders and adversarial training for higher fidelity generation~\cite{kim2021conditional}.
More recently, TTS has moved towards zero-shot voice generation using LLMs and discrete audio representation, such as, Wang et al. \cite{wang2023neural} pioneered this direction through neural codec language modeling with VALL-E, a significantly scaled approach that inspired more efficient variants such as TacoLM by Song et al. \cite{song2024tacolm} and context-aware approaches \cite{xue2024improving}. Current state-of-the-art increasingly combine LLMs with diffusion or autoregressive generative frameworks to improve multilingual capability, and robustness~\cite{tan2024naturalspeech, liao2411fish, liu2024autoregressive} 
Alongside academic advancesm commercial tools are now widely accessible including Google Cloud TTS \cite{googleTTS} and OpenAI’s Audio API \cite{openaiTTS}. Additionally, the open-source community frequently utilizes libraries such as Coqui TTS \cite{coquiTTS} and Tortoise-TTS \cite{tortoise} for local implementation. This not only democratize voice generation 
for creation, but also for exploitation. 

Building on these advances, voice conversion (VC) research has evolved from explicit speaker-content disentanglement to generative frameworks that enable rapid, often zero-shot, speaker adaptation.
For instance, normalization enabled one-shot conversion without speaker-specific retraining
\cite{chou2019one}. Subsequent work improved this separation using contrastive learning to better distinguish speaker attributes~\cite{tang2022avqvc}. Researchers have also developed voice conversion techniques for specific applications, including hierarchical modeling for singing voice conversion \cite{li2022hierarchical}, sequence-to-sequence models for handling emotional expression in speech \cite{yang2022overview}. 

More recently, large language models and diffusion-based generative techniques have driven high-fidelity zero-shot voice conversion~\cite{wang2023lm}. 
Arik et al. proposed Deep Voice 2 demonstrating that multi-speaker TTS models can learn speaker-specific vocal traits~\cite{arik2017deep}. Building on this, SV2TTS uses speaker-verification embeddings to mimic new voices using only a few seconds of audio \cite{jia2018transfer}. 
The field of voice cloning shifted toward zero-shot capabilities afterwards, and Casanova et al. made the first contribution in this regard by developing YourTTS, which integrated VITS with flow-based adaptation for instant voice cloning \cite{casanova2022yourtts}. In recent times, flow-matching and separation techniques have emerged to improve zero-shot techniques further. For instance, Le et al. \cite{le2023voicebox} proposed Voicebox, utilizing non-autoregressive flow-matching for in-context learning, and Qin et al. \cite{qin2023openvoice} introduced OpenVoice to decouple tone color for versatile instant cloning. Moreover, people also widely use commercial tools like ElevenLabs \cite{elevenlabs} and Resemble AI \cite{resembleai} for high-fidelity cloning tasks. Recent advances in open source models have brought state-of-the-art quality to local models~\cite{qwen3tts2026}. 

\subsection{Concerns of Voice Generation}
While automated voice generation tools and techniques have fostered innovation, they have simultaneously introduced many ethical and security concerns that have evolved alongside the technological advancements \cite{brundage2018malicious}. One of the initial challenges regarding this was technical spoofing attacks targeting Automatic Speaker Verification (ASV) systems, where synthetic speech was used to bypass biometric authentication \cite{liu2024autoregressive}. Chen et al. empirically demonstrated this threat, proving that deep learning-based synthesis could successfully bypass widely used authentication systems, such as WeChat \cite{wechat} and Amazon Alexa \cite{alexa}.
As deep learning-based techniques improved fidelity, this threat landscape expanded to more widespread human deception. This contributes to rapid rise of ``audio deepfakes", enabling the deceptive propagation of information convincingly to manipulate people \cite{mirsky2021creation}. In recent times, the availability of zero-shot cloning has resulted in more sophisticated financial fraud and vishing (voice phishing), in which attackers impersonate trusted individuals to authorize large-scale fraudulent transactions \cite{Allen2025RiseAIClonedVoiceScam}. Furthermore, recent studies have indicated that human listeners tend to struggle to reliably distinguish between real and synthesized speech, necessitating the development of robust defense mechanisms to detect synthetic audio like proactive watermarking \cite{san2025human, mai2023warning, zong2025audiomarknet}. Beyond fraud, critical concerns have emerged regarding unconsented usage and copyright of voice data, such as GenAI models increasingly misappropriating the voices of artists and public figures without permission \cite{Zirpoli2025LSB10922}. In this paper, we explore the risks encountered and anticipated by people in voice generation technologies, focusing on both technical vulnerabilities and broader societal implications and how these risks emerge. 

%% file: sections/2-regulation-analysis.tex
\section{Regulatory Environment of Voice Related Generative AI}
\subsection{General US Environment }
    The United States has relatively few comprehensive federal regulations specifically relating to AI compared to other international regulatory bodies. The regulatory environment in the USA is also split, with recent administration signaling a permissive approach, while existing legal trends and lack of protections indicate an unfavorable regulatory environment~\cite{executiveofficeofthepresidentPromotingExportAmerican2025}. The January 2025 Executive Order for Removing Barriers to American Leadership in AI ("Removing Barriers EO")repealed an earlier executive order on AI, which had included measures aimed at mitigating risks from synthetic media. At the same time, the America's AI Action Plan was released ("the Plan"), which identified federal action policies with the aim of securing the US a leadership spot on the global stage of AI~\cite{whitecaseWatchGlobal}. The Plan proposes various incentives to businesses, but the practical impacts on a federal and local level are still unclear. It also states that only ‘unbiased’ LLMs are to be supported by government procurement~\cite{executiveofficeofthepresidentPreventingWokeAI2025}. As unbiased LLMs do not exist, the use of ‘unbiased’ in this context refers to being free from ``ideological dogmas such as DEI", essentially a call for LLMs that are ideologically consistent with the current administration~\cite{whitecaseWatchGlobal}.
    
    The US's regulatory framework regarding AI has historically been decentralized with multiple centers of authority, led by individual states supporting their own initiatives, which can range from focusing on transparency, consumer rights, and algorithmic accountability~\cite{davtyan2025us}. However, the recent administration has communicated to the public that they plan to enact additional executive orders to supersede. 
 Competition and innovation is placed first, potential harms and impacts are framed as a secondary concern. This would place the power squarely on the shoulders of federal regulatory efforts to set the standard for US policy.  
    
\clearpage
\onecolumn
\begin{table*}[t]
\centering
\scriptsize
\setlength{\tabcolsep}{4pt}
\renewcommand{\arraystretch}{1.45}
\rowcolors{2}{RowGray}{white}

\begin{tabular}{p{1.8cm}p{2cm}p{1.5cm}p{1.5cm}p{2.0cm}p{2cm}p{2.25cm}p{2.2cm}}
\toprule
\rowcolor{RowGray}
\textbf{Title} & \textbf{Aim} & \textbf{Voice} & \textbf{Disclosure: who} & \textbf{Disclosure: what} & \textbf{Public} & \textbf{Professional}& \textbf{High-Profile}\\
\midrule

SANDBOX Act[Bill] &
Regulatory sandbox for AI innovation &
\NoIcon 
General AI policy; not voice-specific &
Federal gov agencies, public &
Benefits and risks. &
Public list of participants and potential harms &
Career and income concerns beyond privacy concerns &
Greater advocacy resources may affect waiver renewal\\

SAFE Innovation AI Framework [Memo] &
Federal AI governance framework &
\NoIcon 
Not focused on voice or audio &
\NoIcon &
N/A &
Not a proposed bill or regulation, not applicable. &
Not a proposed bill or regulation, not applicable. &
Not a proposed bill or regulation, not applicable.\\

REAL Political Advertisements Act[Bill] &
Disclosure rules for AI political ads &
\YesIcon 
Requires disclosure of AI-generated audio &
Public &
If the communication has been generated by AI. &
Greater transparency for generated content in political ads. &
May protect political candidates' voice over role
&
Protects against non-consensual political endorsements
\\

Stop Spying Bosses Act[Bill] &
Limit employer surveillance and biometrics &
\YesIcon 
Voice included as biometric data &
Users, public &
Pre-hire surveillance disclosure acknowledgment &
Regulates automated decision, sensitive data collection &
Limits AI AI training data collection contracts  &
Does not have additional impacts \\

AI Research, Innovation \& Accountability Act[Bill] &
Federal oversight and AI research standards &
\YesIcon
Covers AI-generated audio &
Users &
AI-generated content notice requirement&
Platforms Notice when gen AI is used&
Voice actors receive standardize public notice&
Celebrities can influence platform AI usage \\

American Privacy Rights Act[Bill] &
National consumer data privacy law &
\NoIcon
Explicitly excludes audio data &
Users, public &
Privacy policy primary disclosure &
Disclosure and consent required for biometric data &
Easier identification of data brokers &
No specific additional impacts to celebrities\\

No-FAKES Act[Bill] &
Ban unauthorized digital replicas &
\YesIcon
Targets realistic voice/audio clones &
User &
Good Faith removal of non-consensual replicas &
Removal of non-consensual likenesses &
More likely to be the target of digital replicas, provides a pathway for their removal &
More likely to have additional resources to dedicate to removing digital replicas\\

AI Technology Implications[Report]&
Evaluation of Unwanted Robocalls and Robotexts &
\PartIcon
Discusses media and synthetic content &
User &
AI caller disclosure requirement &
Improves robocall transparency and consent &
Voice actors are impacted the same as the general public. &
Celebrities are impacted the same as the general public\\

Consumer Protections for AI[Bill]&
Consumer safeguards for AI systems &
\NoIcon
No explicit voice reference &
User, Deployer &
Foreseeable risks disclosure by Developers &
Personal data explained clearly and accessibly for disabilities &
Voice actors would receive the same notice as the general public &
Celebrities would receive the same notice as the general public\\

AI-Policy Act[Bill] &
National AI strategy and governance &
\YesIcon
Voiceprints defined as biometric data &
User &
That the user is interacting with an AI &
Protecting consumers from AI rights violations &
Voice actors would receive the same notice as the general public &
Celebrities would receive the same notice as the general public\\

Texas Responsible AI Governance Act[Bill] &
State-level AI accountability framework &
\PartIcon
Addresses deepfakes including audio &
User, Government &
Disclosure of AI interaction, annual legislative report &
Protection from government AI identifying specific individuals &
Child sexual impersonation banned in text, not in voice, leaving child voice actors vulnerable &
Child celebrities vulnerable due to text-only restriction\\

CCPA (California)[law]&
Consumer data rights and deletion &
\YesIcon
Voice recordings explicitly covered &
User, Public &
Maintaining the right to know, delete, opt out of the sale of personal information &
Informs people understand personal data use and collectors &
Voice recordings deletable, giving an actors to pursue removal &
Celebrities having same data right as public and same avenue as voice actors\\

Minnesota Consumer Data Privacy Act[Bill] &
Consumer data protection law &
\YesIcon
Voiceprints defined as biometric data &
User &
Disclosure of third parties, copy of user sensitive and organization data &
Consumers to have awareness \& agency over third parties receiving their data &
Voice actors would receive the same rights as the general public &
Celebrities would receive the same rights as the general public \\

Biometric Information Privacy Act[Bill] &
Regulate collection, use, and handling of biometric identifiers &
\YesIcon
Voiceprints defined as biometric data &
User, Public &
Biometric data storage, retention and destruction period &
Informs public on biometric retention and use, with retention not being indefinite &
Voice actors to be more informed of where their voice is being used or stored by private entities &
Celebrities would have the same information as the general public\\

\bottomrule
\end{tabular}

\caption{Comparison of regulations to demonstrate the limitation to address voice generation and synthesis risk}
\label{tab:voice-regulation}
\end{table*}
\twocolumn
\clearpage
    There have been multiple bills proposed to Congress regarding AI regulations, from both political parties and various committees, with some contradictions and also some common patterns across regulations.
    It is essential to note that currently no AI regulatory bill has been passed through Congress~\cite{hr5388congress2025}. 
    All proposed bills (Table~\ref{tab:voice-regulation}) that have been introduced require some level of disclosure about AI practices, though to whom and for what content differ. The most common type of disclosure requirement involves the disclosure of AI practices to the (a) users or customers of the technology, generally regarding the instances of technology use and the collection and extraction of user data. A representative example of an exclusively user-based disclosure regulatory requirement can be found within the Stop Spying Bosses, which requires the instances of workplace surveillance using AI, how it relates to decision making, and what sensitive private data is involved~\cite{caseyStopSpyingBosses2023}. This bill shares a complementary definition of “sensitive private data” with another federal AI bill, the American Privacy Rights Act, and showcasing that the standard for what is not acceptable for AI companies to breach is the barrier of data within this definition~\cite{caseyStopSpyingBosses2023}. 
    
    Another (b) disclosure combination is an overlap in disclosure requirements to both users and the public. 
    This is different from disclosures purely to the users or customers as it requires that AI disclosure notices are accessible even to those not directly affiliated with or impacted by the AI use case, as can be observed in Table 1.  These bills tend to cover more public-impacting AI technologies, such as AI being used for deepfakes that get disseminated publicly~\cite{klobucharRequireExposureAI2023} \cite{noFakesAct2025}. The SANDBOX Act is an exception in disclosure patterns in that it requires disclosure to the federal government in addition to requiring disclosure to the public. The Act shows a different mode of AI accountability that prioritizes tighter relationships between the federal government and AI companies, though not in an antagonistic manner. Rather, the SANDBOX Act allows for the carving out of "obstructive" regulations through waivers \cite{cruzStrengtheningArtificialIntelligence2025}. It is more in alignment with the vision of the Executive branch for AI, in that it is closely related to national security and the continuing position' of the United States as a technology leader. 

    While state AI regulations are no longer being passed and may lose their ability to be enforced to their current capacities, they do serve as models for federal bills and also provide insight into the priorities of localized regions within the USA \cite{duffyTrumpSaysHell2025}. The disclosure categories are mirrored in state and federal bills as showcased in Table 1, though crucially all reviewed state regulations emphasize disclosure to the user of AI practices, even if they also require disclosure elsewhere. User data that is collected, specifically biometric data, is emphasized as something that needs to be disclosed relating to AI technologies to users \cite{cruzStrengtheningArtificialIntelligence2025} \cite{linkBiometricInformationPrivacy2008} \cite{westlinMinnesotaConsumerPrivacy2024}. Transparency and disclosure to the user when they are interacting with or viewing AI is an additional common theme as seen in Table 1. The time of disclosure is more clearly outlined at the state level, as in Texas Responsible AI Governance Act for example, it is explicitly stated to be required before the interaction occurs \cite{capriglioneTexasResponsibleArtificial2025}. Disclosures required for the public also have more specific requirements, stating form of communication, public accessibility, and level of comprehension required for the general public to understand the disclosure. The disparity in specificity is likely to state regulations being existing laws, as opposed to simply proposed bills, with many having years of amendments to ground it in practical applications and shifting priorities, such as the Biometric Information Privacy Act from Illinois, which has existed since 2008 \cite{linkBiometricInformationPrivacy2008}. 

\subsection{General Data Protection Regulation}
    The General Data Protection Regulation (GDPR) is a regulation that was adopted in the European Union (EU) in 2018 and applies to organizations outside the EU if they are handling EU citizen data. The regulation outlines the privacy and security standards required for EU data, levying hefty fines at organizations that violate the GDPR. Broadly, the GDPR covers data protection, accountability, data security, protection by design, consent, the role of data protection officers, and the privacy rights of people \cite{gdprInfo}. The GDPR was created to protect consumers or as referred to in the GDPR, data subjects, from data privacy violations and misappropriations. However, the focus is not solely on privacy policy, as the GDPR has additional requirements for processing that places the burden of responsibility upon data collectors/processors \cite{gdprInfo}. 
    
    A large part of the regulatory power of the GDPR lies in establishing clear definitions of previously ambiguous terms. Personal data is defined as information relating to the identifiable nature of a natural person, or “data subject” \cite{gdprInfo}. In addition the definition of special categories of personal data such as racial or ethnic origin, political opinions, religious or philosophical beliefs, or trade union membership, and the processing of genetic data, biometric data for the purpose of uniquely identifying a natural person, data concerning health or data concerning a natural person’s sex life or sexual orientation, have additional protections \cite{gdprInfo}. Member states that have adopted the GDPR can impose additional protections upon health, genetic, and biometric data. Biometric data is defined as personal data resulting from specific technical processing relating to the physical, physiological or behavioral characteristics of a natural person, which allow or confirm the unique identification of that natural person \cite{gdprInfo}. Within the GDPR the definition for biometric data, voice, is included as a biometric, especially when it relates to their voiceprint for identification.
    
    Data subjects under the GDPR have the right to not be subjected to automated decision making solely, which includes automated profiling \cite{gdprInfo}. The extent of GDPR and AI concludes at data subjects and that is decision making solely that is automated. The EU AI Act is an extension that adds to the obligations of those employing AI systems, namely that AI systems dedicated to interacting with people are required to disclose that they are AI. The exception being those systems that are authorized by law to detect, prevent, or prosecute criminal offenses \cite{EUAIInfo}. 

\subsection{Federal Comparisons}
    Information is to be provided where personal data is collected from the data subject with the contact details of the data protection officer \cite{gdprInfo}. It, like the public disclosure for the SANDBOX Act, is for the public audience. The point of contrast is most divergent in focus of primary disclosure. The SANDBOX Act places the primary focus of disclosure on the disclosure to the federal government, whereas within the GDPR disclosures to governmental agencies are limited to specific scenarios. For example, in scenarios where data disclosure is required to perform a public interest task or it is necessary for the compliance with a legal obligation \cite{gdprInfo} \cite{cruzStrengtheningArtificialIntelligence2025}.
    
    The NO FAKES Act is the primary regulatory attempt in the USA to address the harms of the digital replication of personal likeness through technologies including generative AI. NO FAKES attempts this through creating a standardized definition of a digital replica,an embodied sound, image, or audiovisual recording of likeness of an individual that can be used beyond their contract terms or consent, that can be used to hold organization and platforms liable for using or hosting such replicas \cite{coonsNurtureOriginalsFoster2025}. The GDPR does not explicitly mention digital replicas. However, the provisions of the GDPR can be relevant to the creation of digital replicas. A replication of someone’s voice and likeness without consent or legal exception is a violation of their personal data. In addition, digital replicas without consent can potentially violate the right to be forgotten, right to access data being used, and the right to object to the processing of personal data \cite{gdprInfo}.

%% file: sections/3-method.tex
\section{Research Methods}
\subsection{Data Collection}
\textbf{Data From AI Incident Database}
We analyzed four large-scale crowdsourced and government AI risk incident databases: AI Incident Database (AIID)~\cite{aiid2026}, 
OECD AI Incidents Monitor (AIM)~\cite{oecd_aim_incidents2026}, Federal Trade Commission~\cite{ftc_voice_cloning_challenge2026}, Internet Crime Complaint Center (IC3)~\cite{IC3_website}. We began by applying keyword-based filtering to identify incidents related to the voice generation from the four incident databases. We constructed two keyword groups, one for “voice” contexts (voice keywords) and one for “generation” (generation keyword). In voice keyword group, we considered, voice, audio, speech and in the generation keyword group, we considered clone, deepfake, generation. This initial filtering produced 734 incidents. We then manually reviewed each incident to determine whether it involved voice generation risks. 
Subsequently, researchers came together to discuss, case-by-case, whether each incident was related to certain victim, affected group, and if there is any impact which yielded a collection of 569 incidents. We found four main categories of victims: general public (235), celebrity (112), high-profile individuals (150), voice actors(61~\footnote{there are 29 voice actor overlapping risks included in the celebrity incident since voice actors in those cases were well known celebrities}).

\textbf{Reddit Data}
We collected 2221 Reddit posts and comments using the
Python Reddit API Wrapper (PRAW) from Sept, 2025 to Oct,
2025 (inclusive). 
We conducted iterative manual inspection of Reddit posts during pilot searches. We constructed two keyword groups as we did in the data collection from  AI Incident Database. We conducted open searches across the Reddit platform using combinations of these keywords to ensure comprehensive coverage. Then we removed any duplicates from the search results. 

\textbf{Direct Incident Reported Data}
For risk assessment, relying solely on publicly reported datasets (e.g., incident databases) may overlook critical gaps: such datasets reflect reported issues
but lack direct, unfiltered insights into different group interactions and how these risks occurred. To address this, we expanded our scope by self-curating firsthand incident experienced by people. This ensures access to additional or discovering undocumented ones, unspoken risks, and contextual vulnerabilities that public platforms or retrospective reports might miss. 

We received response from 187 participants totaling, 1067 risk reports in between October 1st to November 15th in the United States from which 96 were voice actors, 71 were general public and 20 were in the category of celebrity (e.g. podcaster, youtuber, internet influencer), high profile actors (e.g.political personnel). This participation in the risk reporting was voluntary. We created a web-based risk reporting tool that would allow participants to submit their experienced and suspected voice generation incidents. We used direct communication with celebrity and high profile individuals or community (National Association of Voice Actors: NAVA)for the risk incident reporting as well as prolific for general public risk reporting. 


\subsection{Data Analysis}
Two researchers conducted an inductive thematic analysis~\cite{braun2006using} across three datasets: Reddit posts, AI incident reports, and direct incident reports. A risk was defined as any scenario or interaction in which voice generation could result in physical, digital, emotional, cognitive, or social harm. Two researchers followed a consistent multi-stage procedure. 
Initially, each independently coded a 20\% sample from each dataset sequentially, beginning with AI incident reports, followed by Reddit posts, and then direct incident reports. This initial round of coding was conducted line by line analysis, with each researcher identifying low-level, fine-grained risk types. These risks are  identified through researchers’ careful observation and analysis of content. 

Following this initial round of coding, we reviewed one another’s labels, discussed any disagreements, and reached an agreement of final coding decisions. If a risk from incidents data aligned with an existing code from the earlier AI incident database coding, we retained the code with supporting examples. If a novel risk emerged, we defined a new code and finalized it through discussion.  Cases in which one researcher identified a risk that others did not, we resolved through consensus. We calculated Cohen’s kappa for the initial 20\% samples across datasets. We calculated inter-rater reliability 0.81. Once we established preliminary codebook, we then divided the remaining 80\% of each dataset among the same two researchers for coding independently. We conducted weekly meetings for consistency and address coding challenges, and refinement. Although the AI incident database provided the initial structure of the codebook, the analysis of AI incident direct reports contributed additional risk types not observed in public forum.  

\subsection{Limitations}
First, voice generation capabilities are evolving and so its' control and defense systems. Thus, the systemic risks may not capture future attack vectors. 
Second, our taxonomy characterizes types of risk and interaction pathways, it does not quantify harm severity, prevalence, or causal mechanisms. We envision this taxonomy as a foundational step toward developing predictive or impact oriented benchmarks.

%% file: sections/4-results.tex
\section{V.O.I.C.E: Risk Taxonomy}
\begin{figure*}
    \centering
    \includegraphics[width=1.1\linewidth]{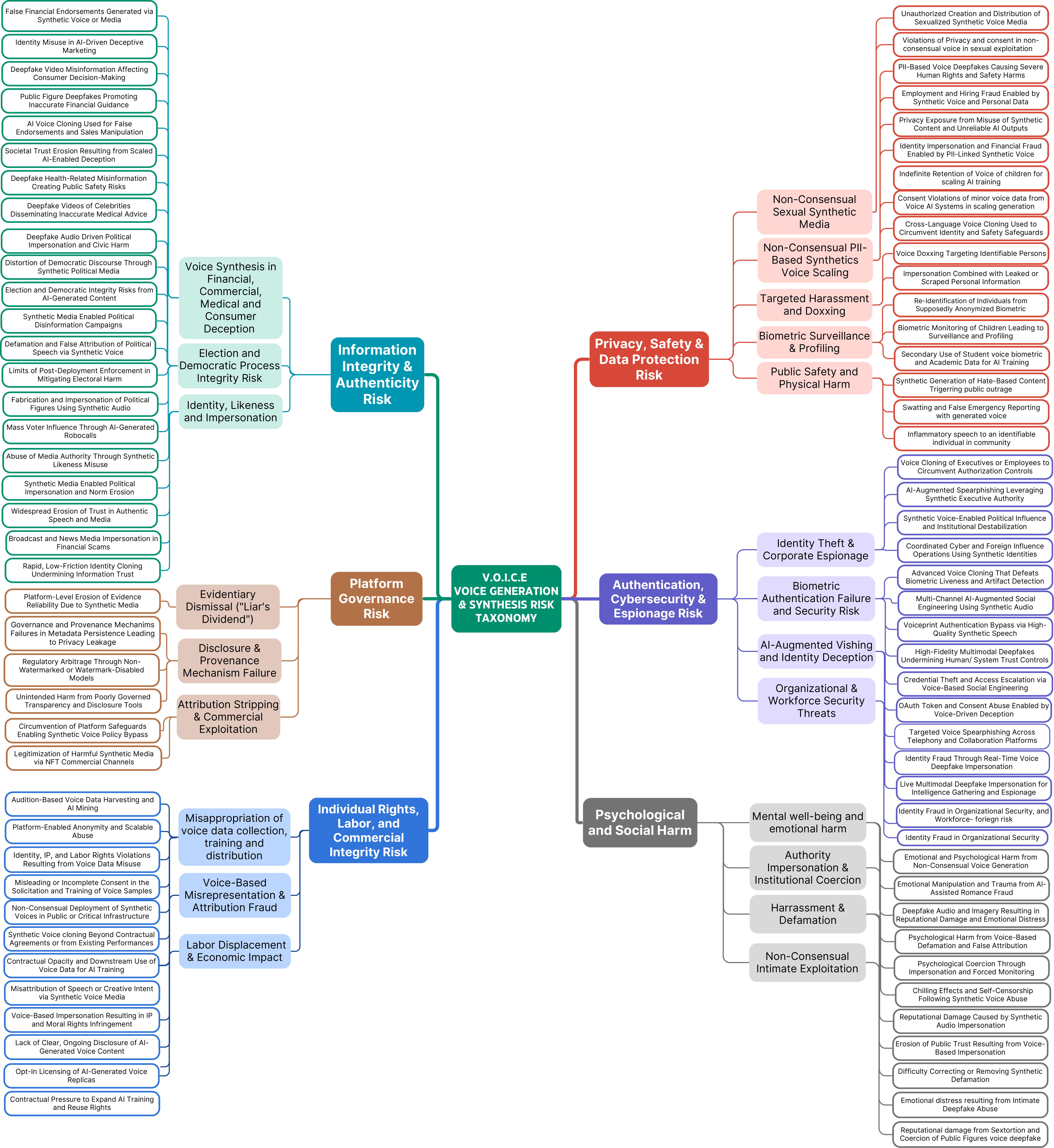}
    \caption{Three-tier taxonomy of Voice Generation \& Synthesis risks. This visualizes the hierarchy, mapping \textbf{six} high-level risk types (First layer) to \textbf{22} medium-level ones (outer ring). \textbf{82} Low-level risks are numbered to align with their corresponding medium-level categories; some names are abbreviated for space. Full taxonomy with definitions and examples is available in the openscience link, which will be updated as new findings and future research refine or expand the identified risk types.}
    \label{fig:risk}
\end{figure*}
To systematically capture the wide range of risks associated with voice  generation technologies, we first identified and labeled low-level risk types across all datasets. These low-level risks represent specific, granular instances of harm, such as ``Re-Identification of Individuals from
Supposedly Anonymized Biometric,'' ``Identity Fraud in Organizational hiring,'' ``Chilling Effects and Self-Censorship
Following Synthetic Voice Abuse,'' ``Reputational damage from Sextortion''. We analyzed each data point, whether a Reddit post, AI incident, or chat log to identify risk patterns, recognizing that a single data point could involve
multiple risk types. Our risk assessments are grounded in Threat modeling~\cite{ai2023artificial}. After identifying all low-level risks, we grouped them into medium and high-level categories, informed by prior AI risk and voice generation and synthesis safety literature (Section~\ref{lit}). This synthesis allowed us to organize related risks into \textbf{six} key high-level types: \textbf{Privacy, Safety \& Data Protection Risk; Authentication, Cybersecurity \& Espionage Risk; Information
Integrity \& Authenticity Risk; Individual Rights, Labor, and
Commercial Risk; Platform Governance Risk; Psychological
and Social Harm} (Figure~\ref{fig:risk}), each representing a distinct domain of risk. The following sections detail each medium- and low-level risk under six high-level risk categories and illustrate their real-world manifestations with definition (Appendix Table~\ref{tab:risk-taxonomy-def}). 
We begin with risk of voice generation and synthesis, their distinct pathways for different affected groups including high profile, voice professional, celebrity and general public. We highlighted findings in cases where one or more affected groups fall into risk categories either because they are directly impacted by the risks or because their voices contributed to (without consent) the risk, an area that is often underexplored in security and privacy literature.

\subsection{Privacy, Safety \& Data Protection Risk}

\begin{mentalcallout}
\textbf{Privacy, Safety \& Data Protection Risk} refers to protection of personal voice data, biometric information, physical safety, and children's privacy, non-consensual PII use, CSAM, and surveillance in the process of voice generation and synthesis.
\end{mentalcallout}

These risks manifest across \textbf{five key dimensions} (Figure~\ref{fig:risk}): public safety threats, non-consensual sexual content, mass PII-enabled fraud, targeted harassment, and biometric surveillance. Children face heightened vulnerability due to their voices being captured in educational settings, smart home devices, and social platforms without adequate consent frameworks or long-term protection guarantees. 

\textbf{Non-Consensual PII- Based Synthetics Voice Scaling.} This risk emerges when voice synthesis is combined with personal identifying information (PII), the risk scales dramatically. This enables mass fraud, targeted exploitation, and severe harms including child sexual abuse material (CSAM)~\cite{yu2025exploring}. Our analysis identifies seven distinct manifestations of this risk, each representing a different vector for exploitation. For instance, in one of the incident occured to \colorbox{blue!15}{\textbf{general public}}, from FBI's Internet Crime Complaint Center (IC3) where \textbf{Child Sexual Abuse Material (CSAM)} were created or modified using generative AI and content manipulation tools which was illegal and being actively prosecuted (from IC3 incident). The public cases show offenders included \textbf{AI-generated voice} with targeted indivduals' other PII (e.g. face, name) to create more convincing content for mass distribution.
We also found intersecting direct incident reports from professionals who repeatedly mentioned their voice being used in pornographics exploitation which was either primarily been collected during audition or for specific roles in games. A direct incident report by \colorbox{green!15}{\textbf{voice actors}} \textit{``In the game Modding, they run my voice through an AI generator, and they've gotten characters to do porn that I did not signed up for. I fear the day when I find my voice to create minor abuse material. Honestly I do not know how to redress it.''} On the same note, an anticipated incident reported by \colorbox{orange!15}{\textbf{youtuber}} \textit{``I have not experienced yet, but I am sure my voice likeness may be misused in future for CSAM endorsement while I am pretty vocal against the corporate like Amazon's misuse of children voice sample and those nasty creators...I have got many anonynous threats on hinting to damage my reputation with the very voice.''}

\textbf{Targeted Harassment and Doxxing.} Voice generation and synthesis amplifies traditional risk by adding a layer of realistic impersonation to doxxing campaigns. Trolls or malicious users took publicly available audio of \colorbox{green!15}{\textbf{voice actors}} (e.g., from podcasts, streams, YouTube, game roles), used ElevenLabs to impersonate real voice actors and publicly post video on Twitter/X that contained personal info such as home addresses, accompanied by racist, offensive, or harassing language, which led the actor to delete their X account (from AI incident database). This incident in particularly indicate not only digital harms, but also loss of autonomy and potential physical harms, amplifying fear and feelings of vulnerability. We found similar direct incident report from many 
\colorbox{cyan!15}{\textbf{high profile individuals}} including individual with political background and internet influencers. As reported \textit{``Five years ago, mostly emails or phone calls with grainy, obviously fake voices. Now, its convincing, doesn’t sound fake at all. My wife panicked several times and my office deal with it every now and then. As a public servant, my voice and face are everywhere on the internet, and that exposure has become a vulnerability. We do not say, but there is always a sense of shame on uncertainties what embarressment might comes up.''}

\textbf{Public Safety and Physical Harm}. This risk emerges when  \colorbox{blue!15}{\textbf{general public}} are both victim and perpetrator where voice cloning to attribute inflammatory, hateful, or discriminatory statements to real individuals, particularly in trusted community settings. For instance, in Baltimore, Maryland, a teacher and athletic director, Dazhon Darien, was arrested after allegedly using AI voice cloning to impersonate a school principal, Eric Eiswert, fabricating racist and antisemitic statements (AI Incident Database). This incident demonstrates how
lowers the barrier for insider threats where perpetrator held a position of access within the institution and exploits community trust. This incident escalate from purely reputational harm to credible physical danger of individuals in leadership positions.

\subsection{Authentication, Cybersecurity \& Espionage Risk} 
\begin{mentalcallout}
\textbf{Authentication, Cybersecurity \& Espionage Risk} refers to
 biometric security failures, corporate espionage, vishing attacks, credential theft, and national security threats through voice synthesis.
\end{mentalcallout}
Across the theme and affected group, there are common concerns on  voice as biometric authentication, a statement from reddit data \textit{``For years, biometrics were the gold standard for authentication. Now, with voice cloning, and synthetic identities on the rise, the very thing meant to make access effortless has become a critical vulnerability. I do not see a way forward. Its now a rabbit hole.''}

\textbf{Identity Theft \& Corporate Espionage.}
Cybersecurity firm Kaspersky reports that criminal and espionage actors are actively deploying AI deepfake videos and voice clones to conduct cyberfraud, identity theft, and corporate espionage (AI Incident Database). High-quality synthetic media is sold on underground markets for \$300–\$20,000 per minute which lower barriers to entry for attackers. This tiered pricing suggests a mature market responding to customer demand (of cyberfraud) where quality, turnaround time, or additional services like accent adaptation or emotional inflection are services. An advanced AI-enabled operation targeted Ben Cardin \colorbox{cyan!15}{\textbf{high profile individuals}}, chair of the U.S. Senate Foreign Relations Committee (AI Incident Database). Attackers impersonated Dmytro Kuleba, a real diplomatic contact known to the senator. After an initial email, attackers conducted a live video call that appeared authentic in appearance and voice, leveraging real-time synthetic video and audio. Suspicion arose only when the impersonator asked politically charged questions related to U.S. elections and foreign policy. 

\textbf{Organizational \& Workforce Security Threats.}  The FBI Internet Crime Complaint Center (IC3) reports a rise in complaints where AI-generated deepfakes (voice and video) combined with stolen PII are used to fraudulently apply for remote positions, notably IT, software, database, and other roles with privileged access to customer PII, financial data, and internal systems (FBI IC3 Report). Observed indicators include voice spoofing during interviews and audio-visual desynchronization, alongside identity theft in pre-employment checks. This represents an enterprise infiltration vector where voice synthesis enables attackers to bypass human verification during remote hiring processes, potentially gaining access to sensitive organizational systems and data. A direct anticipated risk reported to illustrate the other side of the coint \colorbox{blue!15}{\textbf{general public}} \textit{``When I apply for jobs or do remote interviews, I worry if I am talking to a real person as interviewer. I heard of voice harvesting and market for it. A very few known companies that I know will not do it. Now that many companies, often not vetted or very few information, I do worry, but again, I do need a job.''}
\subsection{Individual Rights, Labor, \& Commercial Integrity Risk} 
\begin{mentalcallout}
\textbf{Individual Rights, Labor, \& Commercial Integrity Risk} refers to violations of individual rights, labor protections, intellectual property, and commercial agreements through unauthorized voice use, exploitation of voice professionals, job displacement, and market integrity breaches.
 
\end{mentalcallout}
These risks manifest across \textbf{three key dimensions }misappropriation of voice data for training and distribution, voice-based misrepresentation and attribution fraud, and labor displacement with economic impact and only found among \colorbox{green!15}{\textbf{voice actors}}  and\colorbox{orange!15}{\textbf{celebrity}}  

\textbf{Misappropriation of voice data collection, training and distribution.} Our analysis identifies pathways through which voice professionals' data and performances are exploited without adequate consent, violating privacy right. The Baldur's Gate 3 incident (AI Incident Database) reveals how existing performances for a game, film, or audiobook become potential \textbf{training} data for voice cloning systems. 
Two professional voice actors, Paul Skye Lehrman and Linnea Sage, filed suit against AI startup Lovo, alleging the company illegally \textbf{cloned and commercialized} their voices (AI Incident Database). The actors were misled into providing voice samples under the pretense of a limited project, which were then used without informed consent to create synthetic voices deployed in podcasts and promotional media implicating privacy rights under state law, trademark infringement under the Lanham Act. A scottish voice actor Gayanne Potter alleges that her voice was used without proper consent in AI-generated train announcements \textbf{deployed }by ScotRail (AI Incident Database) in public infrastructure. The case also reveals attribution laundering where passengers hearing announcements in railway do not know they are hearing a synthetic voice, nor do they know the voice is derived from Potter's work. 
To illustrate the risk of voice data training and harvesting, many celebrities and voice professionals shared incidents that they encountered and mediated. As reported by most \colorbox{green!15}{\textbf{voice actors}}  in direct incident \textit{``If an agent sends a very long audition, and the info is very vague, its a direct NO, our community knows it very well, if its not a audiobook job, you DO NOT need to provide more than 60s long audition sample. If its more, they are harvesting voice sample.''} 

\textbf{Labor Displacement \& Economic Impact.}
Another incident reported by a celebrity voice actor, who voiced for a well known game and found the game used AI voice "Pick Up"~\footnote{a “pick-up” is a follow-up recording session where an actor comes back to re-record or add specific lines after the main session is already finished.} in their released version. She reported \textit{``Usual practice is we re-record lines that had mistakes, noise, or technical issues. But this company, they did not hire me for follow up and used AI to clone my voice to add those pick up. This literally replaced my job and even used by voice to clone. quite disgusting. I have a legal proceeding going on about this.''} 
Recently, under a newly negotiated agreement between major game studios and voice actor unions (commonly associated with SAG-AFTRA contracts), voice actors may license AI-generated replicas of their voices to game companies, allowing new voice-overs to be produced without the actor performing them live (AI Incident Database). The arrangement is framed by studios as providing actors with a new revenue stream and by unions as protecting workers through negotiated terms and compensation formulas. However, actors and labor advocates have expressed significant concern that AI voice cloning, even when technically consensual, could devalue human labor, reduce future job opportunities, and shift bargaining power decisively toward studios and platforms. 

\subsection{Information Integrity Risk} 
\begin{mentalcallout}
\textbf{Information Integrity \& Authenticity Risk} refers to
 risk in accuracy, and trustworthiness of generated information in democratic processes, health and safety, consumer deception
\end{mentalcallout}

\textbf{Identity, Likeness and Impersonation.}
Our analysis identifies voice synthesis that enables broader identity impersonation that undermines trust in media, journalism, and public communications. Multiple YouTube channels allegedly used deepfake versions of the voice and likeness of Ravish Kumar, a prominent Indian journalist, to publish fabricated news segments under his name (AI Incident Database). Another large scale impersonation when scammers used AI tools from HeyGen and ElevenLabs to generate deepfake videos of women creators (Michel Janse, Olga Loiek, Shadé Zahrai, Carrie Williams) and misused Lana Smalls's voice, pushing false or offensive products and messages (AI Incident Database). Affected \colorbox{orange!15}{\textbf{celebrity }}and  \colorbox{cyan!15}{\textbf{high profile individual}} discovered their synthetic likenesses promoting products and messages they did not endorse across different platforms and countries (United States, Australia, Ukraine, China, Russia). The multi-platform and multi-national nature of identity impersonation indicates this is not isolated fraud but systematic exploitation of creator identities which creates a governance vacuum where no single entity or authority can effectively address the problem. 

\textbf{Election and Democratic Process Integrity Risk.}
 Our analysis indicate direct threats to democratic legitimacy by enabling voter manipulation, candidate impersonation, and information operations that distort electoral processes. For instance, AI-manipulated audio clips circulated in viral videos on social media falsely depicting Lionel Messi endorsing Argentine presidential candidate Javier Milei ahead of Argentina's national elections (AI Incident Database). This  demonstrates \colorbox{orange!15}{\textbf{celebrity }}  dual harms where they mislead \colorbox{blue!15}{\textbf{general voters }} (direct democratic harm) and damage the impersonated individual's reputation (personal harm).

\subsection{Platform Governance Risk} 

\begin{mentalcallout}
\textbf{Platform Governance Risk} refers to failure in platform policies, disclosure practices, and transparency systems related to inadequate safeguards, attribution stripping, provenance failures, and the "liar's dividend" effect.
\end{mentalcallout}

\textbf{Evidentiary Dismissal ("Liar's Dividend").} Platform governance failure is not the presence of synthetic media but the erosion of trust in authentic evidence that occurs merely because synthetic media exists as a plausible alternative explanation for any recording. From our direct incident reporting, we found many instances by \colorbox{blue!15}{\textbf{general public}}, \colorbox{orange!15}{\textbf{celebrity}}and \colorbox{green!15}{\textbf{voice actors}}. \textit{``Clones and deepfake audio has social injuries on people subjects when they are circulated to viewers who think they are real. It can have a persistent negative impact. Lately, I suspect anything except few legit creators or channels.''} This phenomenon represents epistemic pollution where the information ecosystem's capacity to establish truth deteriorates because now authentic content becomes unverifiable. Traditional evidentiary frameworks in journalism, law, and public accountability have long relied on audio and video recordings as relatively reliable documentation of events~\cite{dubberley2020digital} where voice synthesis fundamentally undermines this evidentiary infrastructure.

\textbf{Disclosure \& Provenance Mechanism Failure.}
 the relationship between provenance mechanisms and privacy creates profound tensions. Robust provenance systems typically require persistent metadata that travels with content across platforms and contexts, documenting who created the content, when and where it was created, what device or software was used, whether it has been edited, and potentially the chain of custody as it moved through different systems. This metadata, while useful for verifying authenticity, can expose creators to serious risks. Journalists documenting government abuses, activists recording human rights violations, whistleblowers leaking corporate malfeasance, or victims documenting domestic abuse may need to prove authenticity of their recordings while protecting their identities, locations, or devices from retaliation. Persistent metadata that enables authentication simultaneously enables identification and targeting. A \colorbox{cyan!15}{\textbf{high profile}} media professional stated \textit{``My main area is political agenda and narrative media is recording, and interviews. I work under a media agency which normally has a political leaning, however, for high profile cases, we do publish under the agency with some protection. I do have concerns when my own and the source [person] identity would be disclosed in provenance.''} This creates a paradox where people who most need to prove authenticity of their recordings are also those who most need to protect their identities. 
 
\textbf{Attribution Stripping \& Commercial Exploitation.} This risk refers to the risk that synthetic voice content is misappropriated and stripped of \colorbox{orange!15}{\textbf{creator (celebrity)}} attribution. 
For instance, an AI-synthetic audio was sold as an NFT on Voiceverse, a marketplace for voice-related digital assets (AI Incident Database). The company later acknowledged that the audio had been created using 15.ai, a free web app for AI voice generation, and was reused without proper attribution or authorization. The case centers on misappropriation and resale of AI-generated content, raising questions about ownership, attribution, and commercialization of synthetic media. This case demonstrates synthetic media laundering that use freely available tools (15.ai) and stripped of any attribution or connection to the original creation context, and commercialized through marketplaces like Voiceverse that provide a veneer of legitimacy through blockchain and NFT infrastructure. 

\subsection{Psychological \& Social Harm}
\begin{mentalcallout}
\textbf{Psychological \& Social Harm} refers to
 emotional, and social well-being impacts from voice synthesis misuse, including non-consensual intimate exploitation, defamation, psychological coercion.
\end{mentalcallout}

\textbf{Authority Impersonation \& Institutional Coercion.} When voice synthesis is used to impersonate authority figures, such as, law enforcement, government officials, institutional representative, it enables psychological coercion that exploits power dynamics and fear of legal consequences among  \colorbox{blue!15}{\textbf{creator (general public)}} particularly immigrants. Criminal actors targeted members of the U.S.-based Chinese community, particularly students, by impersonating Chinese police and diplomatic authorities (FBI Public Service Announcement). The scheme unfolds in four phases: (1) spoofed initial contact posing as trusted institutions; (2) authority escalation with fraudulent credentials; (3) coerced 24/7 video and audio monitoring to control victims and isolate them; and (4) financial extortion. This demonstrates authority exploitation: voice synthesis enables convincing impersonation of specific officials whose authority victims have been socialized to respect and fear.

\textbf{Harassment, Reputational Harm \& Defamation.}  
Voice synthesis amplifies traditional harassment and defamation by adding the persuasive power and emotional impact of synthesized voice, while creating new forms of harm specific to vocal identity appropriation. An AI-generated video used voice synthesis to imitate Erica Lindbeck's voice as Futaba from Persona 5\footnote{a 2016 role-playing video game developed by P-Studio and published by Atlus. The game is the sixth installment in the Persona series}, singing lyrics without her consent (AI Incident Database). Although removed upon request, the incident triggered online harassment. Lindbeck deleted her Twitter account self-censorship under duress showing how voice synthesis creates conditions where victims conclude public platform presence is not worth the psychological cost. Unlike financial losses that can be repaid or technical systems that can be secured, synthetic defamation is challenging to correct or remove.

\subsection{Risk Interaction Pathways}
Based on our empirical evidence, we found six interaction pathaways(Table~\ref{interaction} in Appendix) through which harm can materialize for different stakeholders experiencing distinct vulnerabilities depending on their public exposure and professional context. Pathway 1 (Direct Impersonation) primarily affects high-profile individuals and the general public at the individual level, as attackers target celebrities for reputation damage and ordinary citizens for financial fraud through impersonated phone calls from family members or authority figures. Pathway 2 (Training Data Exploitation) represents a professional-level risk most commonly felt by voice actors whose livelihoods are threatened when their voices collected from audiobook narrations or game performances and repurposed without consent to train AI models that can potentially replace them commercially. This, in turn, create system-level risks through COPPA violations affecting children voice actors whose voices are harvested from platforms and from school records. 
 Pathway 3 (Platform Amplification) operates at both individual and societal levels, disproportionately harm celebrities and politicians through viral deepfakes. Pathway 4 (Biometric Bypass) constitutes a system-level risk and affect high-profile individuals and corporate entities whose voice-authenticated systems become single points of failure, cascading from initial breaches to ransomware and data theft. Pathway 5 (Consent Erosion) represents a societal-level risk where the normalization of unauthorized voice use beginning with seemingly harmless fan content featuring celebrities, gradually undermining consent norms from celebrities to ordinary individuals. Finally, Pathway 6 (Cross-Cutting Interactions) amplifies risks across all levels with the combination of synthesized voices with PII to makes attacks more convincing. 

We also found a range of technical and social engineering tactics in voice generation and how those are propagated leading to risks identified in our taxonomy. Our analysis indicated commercial platforms such as ElevenLabs, Uberduck, Replica Studios, and Jammable provide accessible text-to-speech and voice generation capabilities that have been documented in unauthorized uses ranging from pornographic content creation to celebrity impersonation and defamation.  We also found enterprise and institutional tools including Synthesia, Lovo's Genny platform, ReadSpeaker, and distribution channels like Apple Books AI narration and Spotify AI voice narration demonstrate the mainstreaming of synthetic voice into commercial content production.
The majority of incidents involve "unknown deepfake technology," which indicate a critical challenge where perpetrators leverage anonymous, unregulated platforms (face-swapping apps, NFT marketplaces like Voiceverse and 15.ai) that evade attribution.


%% file: sections/5-discussion.tex
\section{Discussion}

\subsection{Intersectionality \& Differential Risk}
\label{alter1}


Our taxonomy present various type of risks posed by voice generation affecting different group from individual, social and system level. We also found different ways in which the risk propagates through which harm can materialize for different stakeholders with distinct vulnerabilities depending on their public exposure
and professional context. 
For example, high-profile individuals, such as politicians, typically have extensive public visibility, with large volumes of voice data widely available alongside other identifying attributes such as name, face, and biographical details. This parallel with Buolamwini and Gebru's~\cite{buolamwini2018gender} work on how AI systems produce differential harm across demographic group where celebrities and high profile individuals face heightened risk not merely because of their public prominence, but because of auxiliary information, cross platform presence. 
Voice actors similarly have substantial amounts of publicly available voice data through dubbing, advertising, and commercial work (e.g., for virtual assistants or streaming platforms) such as, being voice of siri, alexa, dubbing for netflix, hulu, etc. However, their personal identifying information remains comparatively decoupled compared to high-profile public figures. However, this does not mean they are not affected, rather the risk interaction pathways for voice professionals where this group are exploited with training data exploitation leading to downstream rights violations (\textbf{Pathway 2}) whereas high profile individual may be more affected by direct impersonation leading to immediate harms (\textbf{Pathway 1}).  

This differential risk landscape indicates that  privacy frameworks and regulations may not apply to all subjects. As Stark and Hoffmann~\cite{stark2019data} argue that data collection at scale produces differential exposure, Our findings add nuances where general public, despite having less publicly available data, remains vulnerable to targeted exploitation through adversarial data collection such as scam calls to explicitly harvest voice sample for subsequent fraud (\textbf{Pathway 4}~Table\ref{interaction}). This indicate that low information exposure does not necessarily means low risk when adversaries can actively engineer exposure through deceptive practice. In this work, we develop a taxonomy that characterizes the types of risks emerging from voice generation technologies and the interaction pathways through which these risks manifest across different user groups. As future work, we envision developing impact-oriented benchmarks that operationalize this taxonomy to support the evaluation and design of safer voice AI systems.

Another key finding is power asymmetries that shape disparities in access to regulatory resources and protections. This observation parallels Cohen’s analysis of contemporary information economies~\cite{cohen2019between}, which argues that complex legal and technical infrastructures often systematically privilege actors with greater resources, institutional support. Our regulatory analysis indicates that high profile individuals have some advocacy and protection against AI clones compared to the general public. Also, celebrities can deploy legal teams to pursue take-down request and leverage media relationship where we observed voice actors to self censor themselves after being harassed by non-consensual imagery. Similarly, general public have fragmented platform reporting systems to effectively dispute impersonations. This asymmetry means that voice generation technologies operationalize and amplify existing inequalities.  
\subsection{Implication of V.O.I.C.E Taxonomy}

\textbf{Tiered Regulation.} Voice risk is not uniform, it varies with information exposure, interaction pathways and resources availability for affected group to redress. This implies the regulatory unit should not only be the model specific, but also the use case, target group, risk pathway. Our incident analysis reveals an exposure-based risk dimension in which harm scales with quantifiable factors, hours of publicly available audio, degree of cross-platform replication, and the presence of aligned identity metadata such as names, images, and biographical information \cite{muller2022does, sharma2025aligning}. However, exposure alone does not determine societal harm. Risk also depends on downstream propagation dynamics how synthetic content spreads. 
Thus, we recommend regulatory bodies to adopt tiered safety frameworks that combine exposure metrics with propagation factors, imposing graduated requirements and penalties based on aggregate risk scores. This approach operationalizes "imposing costs" \cite{moore2010economics, sharma2020analysis}, entities deploying high-risk voice synthesis face proportionately higher compliance burdens and penalties for violations, creating economic incentives aligned with harm prevention.

\textbf{Safety Modeling for Platform Governance.} A practical policy move is to internalize externalities, if voice generation products create measurable profit, they should also carry measurable responsibility funding. Specifically, providers of voice generation services to contribute a fixed share of voice-related revenue, or usage-based fees scaled by risk tier, to a dedicated fund supporting  shared provenance infrastructure and independent remediation and rapid-response mechanisms. 
For services that enable realistic voice cloning, may establish escrowed funds that are released only if providers meet audit benchmarks (watermark robustness, complaint handling SLA, transparency reports). This mechanism shifts the cost of enforcement from affected individuals to providers, who are best positioned to prevent and remediate harm.

\section{Acknowledgement}
We thank our participants for their contributions and sharing their
insights. We thank National Association of Voice Actors (NAVA) and 	
Tim Friedlander NAVA Founder and President for making introduction of our researchers to the community during our initial research with voice professionals. We also thank our fellow researchers who made connection to different group of participants, including, celebrity, internet influencer, podcaster which made this research possible.  

%% file: sections/Appendix.tex
\section{Appendix}
\begin{table*}[t]
\centering
\scriptsize

\begin{tabularx}{\linewidth}{p{3cm} X p{4cm} p{4cm}}
\hline
 & \textbf{Stage} & \textbf{Enabling Factors} & \textbf{Example} \\
\hline

\rowcolor{RowGray} \textbf{Pathway 1: Direct Impersonation → Immediate Harm} & Voice synthesis used to impersonate specific individuals for immediate malicious objectives& &\\
 \hline
 
 &\textit{Voice Acquisition} & Attacker obtains voice samples from public sources (interviews, social media, performances)& Politician speeches, podcast recordings, actor performances
 \\
  &\textit{Synthesis} & Accessible consumer-grade tools, low technical barrier& ElevenLabs, Uberduck, unknown tools, etc
 \\

    &\textit{Contextualization} & Synthetic voice combined with context for credibility& PII, voice (family member fraud), authority signals (police impersonation)
 \\
   &\textit{Deployment} & Content delivered via channels maximizing impact& Phone calls, robocalls, video platforms, messaging apps
 \\
    &\textit{Harm Realization} & Temporal lag in detection, verification difficulty, reactive enforcement & Financial loss, voter suppression, emotional distress, safety threats
 \\

\rowcolor{RowGray} \textbf{Pathway 2: Training Data Exploitation → Downstream Rights Violations} & Voice data collected for one purpose, repurposed for AI training, enabling systematic exploitation
& &
 \\
   &\textit{Initial collection} & Contractual ambiguity, COPPA violations, educational/service contexts &Audiobook narration, game performance, Alexa interactions, student recordings 
 \\
    &\textit{Repurposing} &Data used for AI model training without clear consent & Apple/Spotify narrator voices, EA Apex Legends, Amazon child voice data, University of Michigan
 \\

    &\textit{Model Deployment} &Trained models enable synthetic voice generation & Replica Studios, Apple Books AI narration, commercial voice synthesis
 \\
     &\textit{Collective Impact} &Economic incentives favor replacement & Voice acting profession threatened, children face lifetime vulnerability
 \\

\rowcolor{RowGray} \textbf{Pathway 3: Platform Amplification → Viral Misinformation} & Synthetic voice content achieves viral spread through platform mechanisms before verification& &
\\
     &\textit{Content Creation} &Accessible synthesis tools, minimal detection at creation & Celebrity endorsement frauds, political deepfakes, medical misinformation
 \\

      &\textit{Initial Posting} &Platform upload permissiveness, no mandatory disclosure & TikTok, X, YouTube, Facebook, Instagram, WhatsApp
 \\

       &\textit{Algorithmic Amplification} &Engagement optimization, cross-posting&  Viral deepfakes (Elon Musk Grupo Globo, Joe Rogan supplements)
 \\
       &\textit{Detection Lag} &Verification occurs after significant exposure&  Dr. Jim Mann metformin case, government facility crisis image
 \\
   &\textit{Incomplete Remediation} &Removal on some platforms, persistence on others& Content takedowns while mirrors/downloads persist
 \\

\rowcolor{RowGray} \textbf{Pathway 4: Biometric Bypass → Security Cascade} & Voice synthesis defeats authentication, enabling access that cascades to broader breaches& &
\\
 &\textit{Authentication Targeting} &Voice as single-factor authentication, inadequate liveness detection& Bank voice authentication, corporate VPN access, government services
 \\
  &\textit{Privilege Escalation} & Trust assumptions post-authentication, lateral movement capability & Credential theft, data exfiltration, ransomware deployment
 \\
  &\textit{Persistent Access} & Attackers establish ongoing compromise & Black Basta ransomware, corporate espionage, data theft
 \\
   &\textit{Downstream Impacts} & Cascading failures from initial breach & Financial loss, intellectual property theft, ransomware
 \\

\rowcolor{RowGray} \textbf{Pathway 5: Consent Erosion → Normalized Exploitation} & Gradual normalization of voice synthesis without consent through incremental boundary violations& &
 \\
   &\textit{Initial "Harmless" Uses} & Synthesis framed as fan content, parody, homage & Skyrim modding (initially non-sexual), character singing
 \\
    &\textit{Commercial Exploitation} & Market infrastructure (NFT platforms), attribution stripping &  unauthorized voice useNFT sales (Voiceverse/15.ai), commercial mod distribution
 \\
   &\textit{Normalized Violation} & Consent becomes exception rather than rule &  Default assumption that public voices are "fair game"
 \\
 
 \rowcolor{RowGray} \textbf{Pathway 6: Cross-Cutting Interaction Patterns} & Synthetic voice combined with personally identifiable information creates compound credibility & Each element individually creates certain level of credibility & PII \& Voice Synthesis Convergence

 \\


\hline

\end{tabularx}
\caption{Primary Interaction Pathways for Voice Generation Harms from Empirical Data}
\label{interaction}
\end{table*}

\clearpage
\onecolumn
\begin{table}[t]
\centering
\scriptsize
\renewcommand{\arraystretch}{1.2}
\begin{tabularx}{\textwidth}{p{0.25\textwidth} |p{0.24\textwidth} |>{\RaggedRight\arraybackslash}X}
\toprule
\textbf{High Level Risks} & \textbf{Medium Level Risks} & \textbf{Medium Level Risk Definition} \\
\midrule

\multirow{2}{*}{\textbf{Privacy, Safety \& Data Protection Risk}}
& \textbf{Non-Consensual Sexual Synthetic Media}
& AI voice cloning systems are exploited to create and distribute pornographic or sexually explicit content that falsely represents real individuals 
\\
\cmidrule(lr){2-3}
& \textbf{Non-Consensual PII-Based Synth Voice Scaling}
& PII is combined with AI-generated synthetic voice technologies to enable scaled impersonation, identity theft, cross-language circumvention of safeguards causing privacy and consent violations
\\
\cmidrule(lr){2-3}
& \textbf{Targeted Harassment and Doxxing}
&  AI Synthetic voice deployed to impersonate, intimidate, or facilitate the exposure of private information (doxxing) of identifiable individuals, combining voice cloning with leaked or scraped personal data\\
\cmidrule(lr){2-3}
& \textbf{Biometric Surveillance \& Profiling}
&  Voice data, as a form of biometric information, is collected, stored, and repurposed without informed consent from vulnerable populations 
\\
\cmidrule(lr){2-3}
& \textbf{Public Safety and Physical Harm}
& AI-generated synthetic voice content is weaponized to trigger real-world violence, emergency response manipulation, through false attribution   \\
\midrule

\multirow{2}{*}{\textbf{Cybersecurity \& Espionage Risk}}
& \textbf{Identity Theft \& Corporate Espionage}
& High-fidelity AI voice cloning is used to impersonate executives, employees, or trusted organizational figures to bypass authorization controls\\
\cmidrule(lr){2-3}
& \textbf{Biometric Authentication Failure}
& Risk in voice-based biometric authentication systems including liveness detection and artifact by replicating legitimate users' voices \\
\cmidrule(lr){2-3}
& \textbf{AI-Augmented Vishing \& Identity Deception}
& Social engineering attacks (vishing) with credibility, and scale of impersonation and enable attackers to harvest credentials, manipulate OAuth, gain network access, real-time multimodal deepfake \\
\cmidrule(lr){2-3}
& \textbf{Organizational \& Workforce Risk}
& Synthetic audio and video to perpetrate identity fraud in hiring, employment verification, and onboarding processes resulting in data breach\\
\midrule

\multirow{3}{*}{\textbf{Individual Rights, \& Labor Risk}}
& \textbf{Misappropriation of voice collection, training \& distribution}
& Risk that AI systems train on, replicate, or deploy identifiable individuals' voices through misleading consent processes, audition data harvesting, contractual opacity\\
\cmidrule(lr){2-3}
& \textbf{Voice-Based Misrepresentation \& Attribution Fraud}
& AI-generated synthetic audio falsely attributes speech, creative intent, or artistic expression to identifiable individuals without authorization \\
\cmidrule(lr){2-3}
& \textbf{Labor Displacement \& Economic Impact}
& AI-generated synthetic voice contribute to systematic devaluation of labor, large-scale economic displacement, erosion of workers' bargaining power \\
\midrule

\multirow{2}{*}{\textbf{Information Integrity \& Authenticity Risk}}
& \textbf{Voice Synthesis in Financial, Commercial, Medical \& Consumer Deception}
&  AI-generated synthetic voice is to falsely depict endorsements, manipulate consumer decisions, misrepresent product safety or efficacy \\
\cmidrule(lr){2-3}
& \textbf{Election \& Democratic Process Integrity Risk}
& Risk that AI-generated synthetic audio is weaponized to impersonate public officials, manipulate voter perceptions, distort democratic discourse\\
& \textbf{Identity, Likeness \& Impersonation}
& Risk arise from low-friction identity cloning with voice to impersonate real individuals, including journalists, public figures, and private citizens \\
\midrule

\multirow{3}{*}{\textbf{Platform Governance Risk}}
& \textbf{Evidentiary Dismissal ("Liar's Dividend")}
&  Existence and plausibility of AI-generated synthetic media undermines trust in authentic recordings, enabling bad actors to dismiss genuine evidence\\
\cmidrule(lr){2-3}
& \textbf{Disclosure \& Provenance Mechanism Failur}
& Risks that reliance on watermarking, content authentication, or AI-disclosure mechanisms and from privacy leakage metadata tampering\\
\cmidrule(lr){2-3}
& \textbf{Attribution Stripping \& Commercial Exploitation}
& Risks that AI-generated synthetic voice is stripped of creator attribution or resold through digital marketplaces (NFTs and licensing platforms)\\
\midrule
\multirow{3}{*}{\textbf{Psychological \& Social Harm}}
& \textbf{Mental well-being \& emotional harm}
&  AI-assisted synthetic voice media is used to inflict psychological distress, emotional manipulation, or trauma through romance fraud, defamation\\
\cmidrule(lr){2-3}
& \textbf{Authority Impersonation \& Institutional Coercion}
& Risks enabling impersonation of authority figures, including law enforcement or institutional leaders to psychologically coerce victims\\
\cmidrule(lr){2-3}
& \textbf{Harrassment \& Defamation}
& AI-generated synthetic voice media is weaponized to impersonate individuals for purposes of online harassment\\
\cmidrule(lr){2-3}
& \textbf{Non-Consensual Intimate Exploitation}
& Harm manifests as profound reputational damage, psychological trauma, emotional distress, social consequences, sexual exploitation\\
\bottomrule
\end{tabularx}
\caption{Hierarchical structure of V.O.I.C.E risk taxonomy: high-level and medium-level risks Definition}
\label{tab:risk-taxonomy-def}
\end{table}
\twocolumn
\clearpage